\documentclass[conference, 10pt]{IEEEtran}
\usepackage{cite,graphicx,subcaption}
\usepackage{color}
\usepackage{placeins}
\usepackage{float}
\usepackage{tabularx,colortbl,booktabs}
\usepackage{amsmath,amsthm,amssymb,pifont}
\usepackage{dsfont}

\begin{document}
\title{ANALYSIS AND PREDICTION OF JND-BASED VIDEO QUALITY MODEL}

\author{
\IEEEauthorblockN{Haiqiang Wang, Xinfeng Zhang, Chao Yang and C.-C. Jay Kuo}
\IEEEauthorblockA{University of Southern California, Los Angeles, California, USA\\
\{haiqianw, xinfengz, yangchao\}@usc.edu, cckuo@sipi.usc.edu\\}
}

\maketitle

\begin{abstract}
The just-noticeable-difference (JND) visual perception property has
received much attention in characterizing human subjective viewing
experience of compressed video. In this work, we quantify the JND-based
video quality assessment model using the satisfied user ratio (SUR)
curve, and show that the SUR model can be greatly simplified since the
JND points of multiple subjects for the same content in the VideoSet can
be well modeled by the normal distribution.  Then, we design an SUR
prediction method with video quality degradation features and masking
features and use them to predict the first, second and the third JND
points and their corresponding SUR curves.  Finally, we verify the
performance of the proposed SUR prediction method with different
configurations on the VideoSet. The experimental results demonstrate
that the proposed SUR prediction method achieves good performance in
various resolutions with the mean absolute error (MAE) of the SUR
smaller than 0.05 on average.
\end{abstract}


\section{Introduction}

Real-time video streaming contributed to the largest amount of Internet
traffic in both fixed and mobile networks according to the global
Internet phenomena report in 2016 \cite{sandvine2016global}. It
accounted for 71\% downstream bytes of fixed access and 40\% of
downstream bytes of mobile access, respectively. The streaming service
providers, such as Netflix, YouTube, and Amazon, strive to provide users
with the best viewing experience given the constraints of network
bandwidth and viewing device resolutions. For example, a Per-Title
encode optimization technique \cite{per_title_encode} was proposed to
run analysis on an individual content to determine the optimal encoding
recipe based on its complexity. However, due to diversity of the video
content and display devices, it is a very challenging problem to develop
a Video Quality Metric (VQM) that can accurately and consistently
measure human perceptual quality of a video stream.

In the past decades, there has been a large amount of efforts in
developing new visual quality metrics to address this problem, including
SSIM \cite{ssim}, FSIM \cite{fsim}, DLM \cite{dlm}, etc. These methods
have been evaluated on several public databases and achieve high
consistency with the calibrated scores by human. In the existing
databases, the distorted images and videos are assigned a set of
discrete or continuous values called opinion scores; and the typical
opinion scores in the range [1,5], with 5 being the best quality and 1
denoting the worst quality. The quality of the images/videos are
acquired by averaging these subjects' opinion scores, named mean opinion
scores (MOS). However, there is one shortcoming with these calibrated
databases. That is, the difference of selected contents for ranking is
sufficiently obvious for a great majority of subjects and easy to
distinguish.

In fact, humans cannot perceive small pixel variation in coded
images/videos until the difference reaches a certain level, which is
denoted as just-noticeable-difference (JND). There is a recent trend to
measure the JND threshold directly for each individual subject. The idea
was first proposed in \cite{lin2015experimental}. An assessor is asked
to compare a pair of coded image/video contents and determine whether
they are the same or not in the subjective test, and a bisection search
is adopted to reduce the number of comparisons. Two small-scale
JND-based image/video quality datasets were built by the Media
Communications Lab at the University of Southern California,
\textit{i.e.}, MCL-JCI dataset \cite{jin2016statistical} and MCL-JCV
dataset \cite{mcl_jcv}. They target at the JND measurement of JPEG coded
images and H.264/AVC coded videos, respectively. More recently, a
large-scale JND-based video quality dataset, called the VideoSet, was
built and reported in \cite{Wang2017292}. The VideoSet consists of 220
5-second sequences, each at four resolutions (i.e., $1920 \times 1080$,
$1280 \times 720$, $960 \times 540$ and $640 \times 360$). Each of these
880 video clips was encoded by the x264 encoder implementation
\cite{aimar2005x264} of the H.264/AVC standard with $QP=1, \cdots, 51$
and the first three JND points were evaluated by 30+ subjects.

The JND reflects the boundary of perceived quality changes, which is
well suitable to determine the optimal image/video quality with minimum
bit rates. There are also several JND prediction methods based on the
these JND-based datasets proposed in literatures.  Huang \textit{et al.}
\cite{huang2017measure} proposed a JND prediction method by utilizing
the masking effect related features, and designed a spatial-temporal
sensitive map (STSM) to capture the unique characteristics of the source
content. In \cite{wang2017prediction}, a machine learning framework was
proposed to predict the Satisfied User Ratio (SUR) curve of the first
JND. It took both the local quality degradation as well as the masking
effect into consideration and extracted a compact feature vector to feed
it into the support vector regressor to obtain the predicted SUR curve.

In this paper, we revisit the JND modeling problem with an in-depth
analysis on practical compressed video. Based on the analysis, we
propose to model the JND samples of multiple subjects by normal
distribution. Furthermore, we extend our previous work in
\cite{wang2017prediction} to predict the second and the third JND points
under the normal distribution assumption of JND samples. We validate the
proposed JND point prediction method on the VideoSet under different
test settings, and conclude that the proposed method can well predict
the JND points. On the other hand, the prediction accuracy is sensitive to the QP value of
anchor clips. Finally, we show possible applications of the proposed
JND/SUR model in adaptive video streaming services.

\begin{figure*}[!t]
\centering
	\begin{subfigure}[b]{1.0\linewidth}
	\centering{}
	\includegraphics[width=0.19\linewidth]{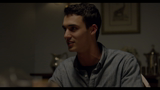}
	\includegraphics[width=0.19\linewidth]{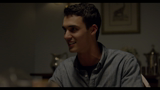}
	\includegraphics[width=0.19\linewidth]{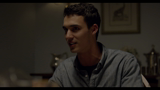}
	\includegraphics[width=0.19\linewidth]{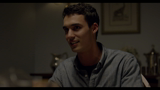}
	\includegraphics[width=0.19\linewidth]{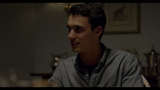}
	\caption{\label{}}
	\end{subfigure}
  \begin{subfigure}[b]{1.0\linewidth}
	\centering
	\includegraphics[width=0.19\linewidth]{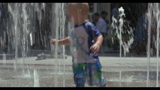}
	\includegraphics[width=0.19\linewidth]{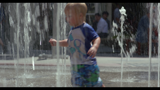}
	\includegraphics[width=0.19\linewidth]{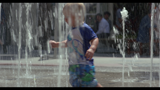}
	\includegraphics[width=0.19\linewidth]{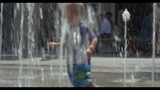}
	\includegraphics[width=0.19\linewidth]{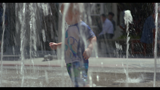}
	\caption{\label{}}
	\end{subfigure}
\caption{Representative frames from source sequences: (a) \#37
and (b) \#90.} \label{fig:frames_of_two_sequences}
\end{figure*}

The rest of the paper is organized as follows. The JND and SUR curve
models are introduced in Sec. \ref{sec:JND_SUR_model}. The proposed SUR
prediction method for multiple JND points is introduced in Sec.
\ref{sec:sur_pred}. Experimental results are provided in Sec.
\ref{experiment_results}. Finally, concluding remarks and future
research direction are given in Sec. \ref{sec:conclusion}.

\section{Just-Noticeable-Difference (JND) and Satisfied User Ratio
(SUR) Modeling}\label{sec:JND_SUR_model}

For a source video clip denoted by $r$, it can be compressed into a set
of clips $d_{i}$, $i=0, 1, 2, \cdots, 51$, where $i$ is the quantization
parameter (QP) index used in H.264/AVC. Typically, clip $d_{i}$ has a
higher PSNR value than clip $d_{j}$, if $i<j$, and $d_{0}$ is the
losslessly coded copy of $r$. The JND of coded clips characterizes the
distortion visibility threshold with respect to a given anchor, $d_{i}$.
Through the subjective experiment, JND points can be obtained from a
sequence of consecutive Noticeable/Unnoticeable difference tests between
clips pair $(d_{i}, d_{j})$.

The anchor, $d_{i}$, is fixed while searching for the JND location. We
adopt a binary search procedure and it takes at maximum $L=6$ rounds to
find the JND location.  We use $l$, $l = 1, \cdots, L$, to indicate the
round number and $m$, $m = 1, \cdots, M$, to indicate the subject index,
respectively. The result is denoted by a random variable $X_{m, l}$.  If a
pair has noticeable difference, we set $X_{m, l}=1$. Otherwise, $X_{m,
l}=0$.
The JND test outcome obtained from the $m^{th}$ subject
at the $l^{th}$ round is
\begin{equation}\label{eq:p}
Pr(X_{m, l}=1) = p_{m, l},
\end{equation}
where $p_{m, l}\in [0,1]$ indicates the confidence of making the ``noticeable difference'' decision of the $m^{th}$ subject at the $l^{th}$ round.

The probability $p_{m, l}$ is close to one for smaller $l$ since the
quality difference between two clips in comparison is clearer in early
test rounds. It is close to 0.5 for larger $l$ as the coded clip
approaches the final JND location.  It is assumed that $p_{m, l}$ is
identically distributed for a specific $l$.  Meanwhile, $X_{m, l}$ is
independently distributed because subjects are randomly recruited. The
JND interval at round $l$, denoted by $\Delta QP_l$, can be expressed as
\begin{equation}
\Delta QP_l= \Delta QP_0 (\frac{1}{2})^l,
\end{equation}
where $\Delta QP_0=51$ is the initial JND interval.
Finally, the JND location for subject $m$ can be written as
\begin{equation} \label{eq:decomposition}
Y_m = \sum_{l=1}^{L} (1-X_{m,l}) \Delta QP_l,
\end{equation}
since we need to add $\Delta QP_l$ to the offset (or left) point
of the current JND interval if $X_{m,l}=0$ and keep the same offset
if $X_{m,l}=1$.
Then, Eq. (\ref{eq:decomposition}) can be simplified to
\begin{equation} \label{eq:decomposition}
Y_m = \sum_{l=1}^{L} \Delta QP_l - \Delta QP_0 \sum_{l=1}^{L} X_{m,l}
(\frac{1}{2})^l,
\end{equation}
where the first term is a constant. We are not able to find a closed
form for the second term. For this reason, we plan to study the
relationship between this term and the probability model in Eq.
(\ref{eq:p}) by numerical simulation in the future.


On the other hand, we have verified the JND normal distribution
assumption by the standard normality test \cite{jarque1987test} in Table
\ref{tab:normality_check}. We see from the table that there is a high
probability for the measured JND samples to pass the standard normality
test.

\begin{table}[!t]
	\scriptsize
\centering
\caption{The percentages of JND samples that pass the normality test,
where the total sequence number is 220.}\label{tab:normality_check}
\vspace{3mm}
\begin{tabular}{*{4}{c}}
\toprule
Resolution          & The first JND & The second JND & The third JND\\ \midrule
    1080p           & $95.9\%$       & $95.9\%$       &  $93.2\%$     \\
    720p            & $94.1\%$       & $98.2\%$       &  $95.9\%$     \\
    540p            & $94.5\%$       & $97.7\%$       &  $96.4\%$     \\
    360p            & $95.9\%$       & $97.7\%$       &  $95.5\%$     \\
    \bottomrule
\end{tabular}
\end{table}

It is worth mentioning that the JND is flexible in modeling the
subjective viewing experience because the anchor clip can be any coded
clip. For simplicity, we use the first JND to demonstrate the
methodology to predict the QoE. The same methodology applies to other
JND points as well. The first JND location is the transitional index $j$
that lies on the boundary of perceptually lossless and lossy visual
experience for a subject. We first transfer the QoE measure to the
satisfied user ratio (SUR) prediction. A viewer is satisfied for a
compressed video if it appears to be perceptually the same as its
reference/anchor.  Mathematically, the satisfied user ratio (SUR) of
video clip $d_{j}$ can be expressed as
\begin{equation}\label{eq:sur_sum}
S_{j} = 1-\frac{1}{M}\sum_{m=1}^{M}\mathds{1}_{m}(d_{j}),
\end{equation}
where $M$ is the total number of subjects and $\mathds{1}_{m}(d_{i})=1$
is the indicator function. $\mathds{1}_{m}(d_{i})=1$ or $0$ if the $m^{th}$
subject can or cannot see the difference between compressed clip $d_{i}$ and
its reference, respectively. The summation term in the right-hand-side of
Eq. (\ref{eq:sur_sum}) is the empirical cumulative distribution function (CDF)
of JND samples. Then, we can obtain a compact formula for the SUR curve as
\begin{equation}
S_{j} = Q(d_{j}|\mu, \sigma^{2}),
\end{equation}
where $Q(\cdot)$ is the Q-function of a Normal distribution. $\mu$ and $\sigma^{2}$ are the sample mean and variance of JND samples, respectively.

\begin{figure*}[!htb]
\centering
\begin{subfigure}[b]{0.3\linewidth}
\centering
\includegraphics[width=1.0\linewidth]{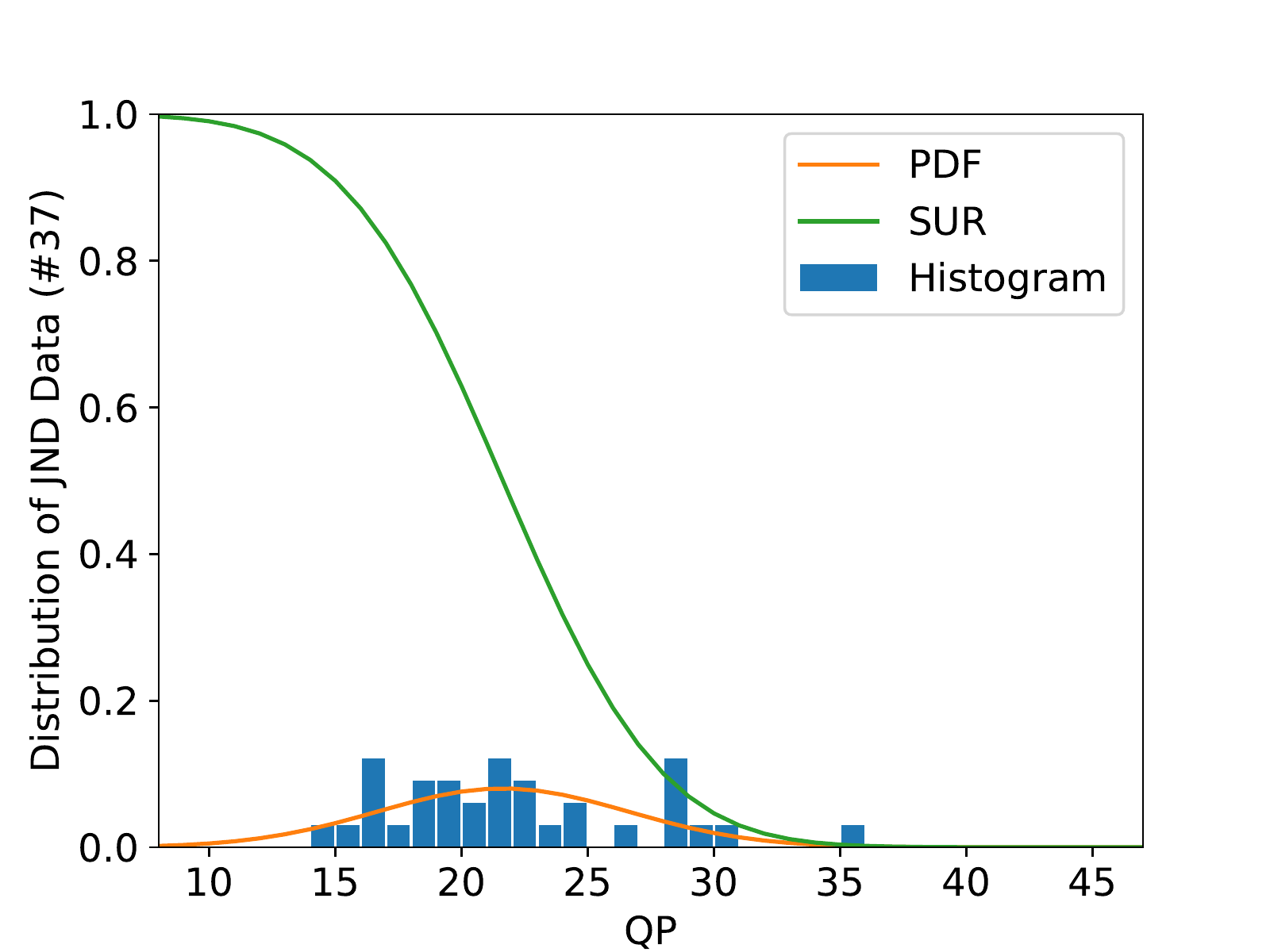}
\subcaption{}
\end{subfigure}
\begin{subfigure}[b]{0.3\linewidth}
\centering
\includegraphics[width=1.0\linewidth]{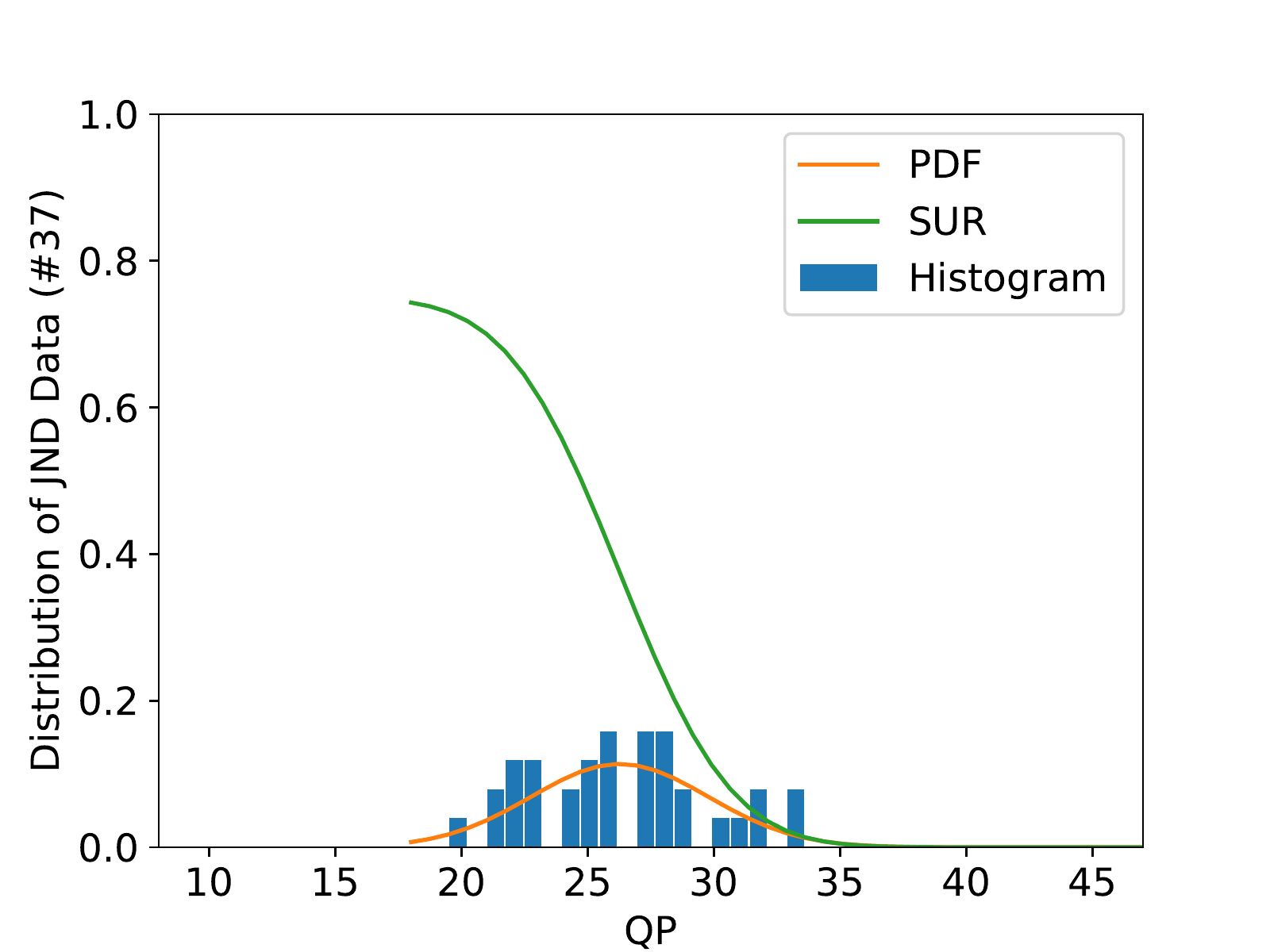}
\subcaption{}
\end{subfigure}
\begin{subfigure}[b]{0.3\linewidth}
\centering
\includegraphics[width=1.0\linewidth]{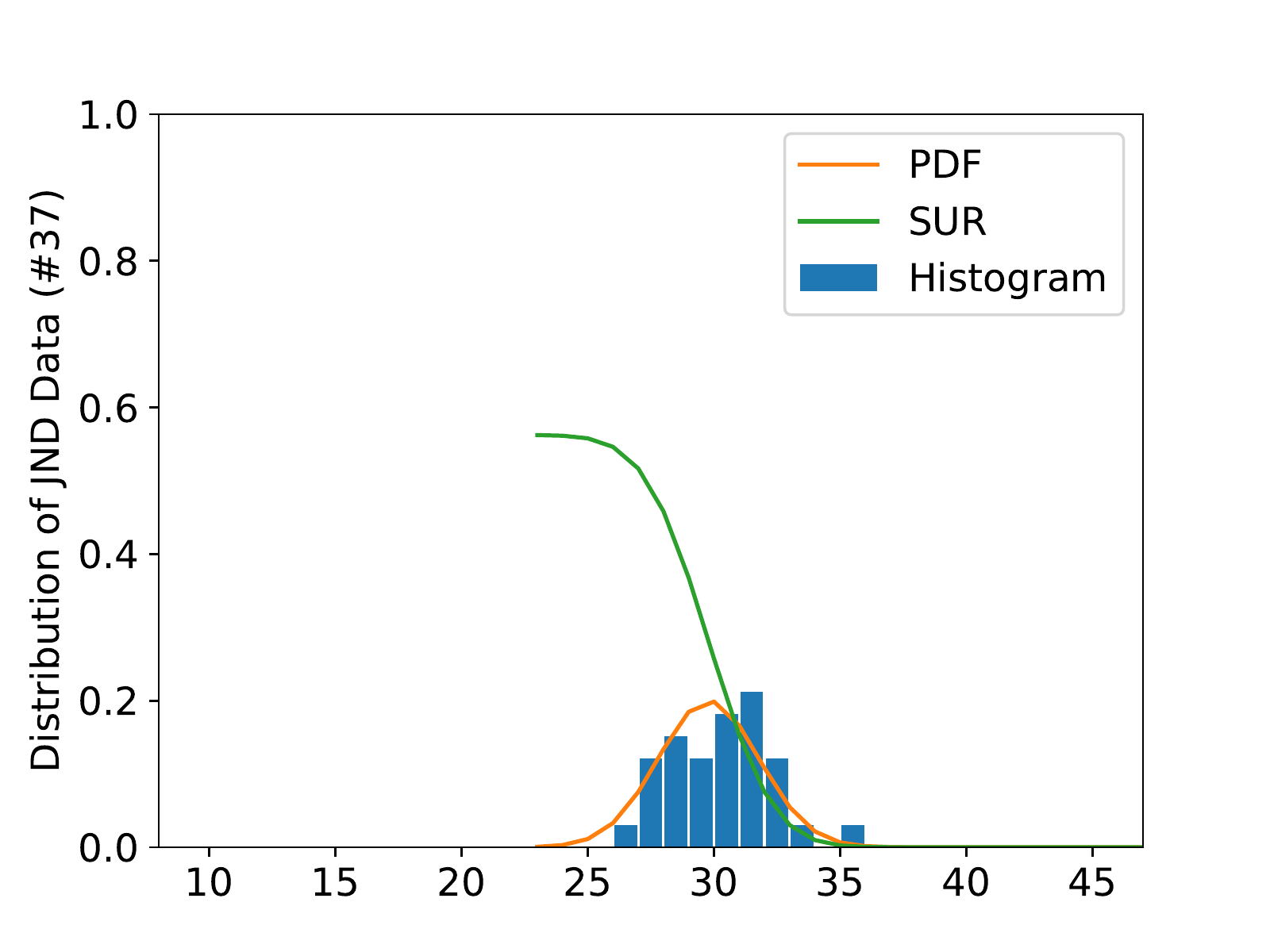}
\subcaption{}
\end{subfigure}
\begin{subfigure}[b]{0.3\linewidth}
\centering
\includegraphics[width=1.0\linewidth]{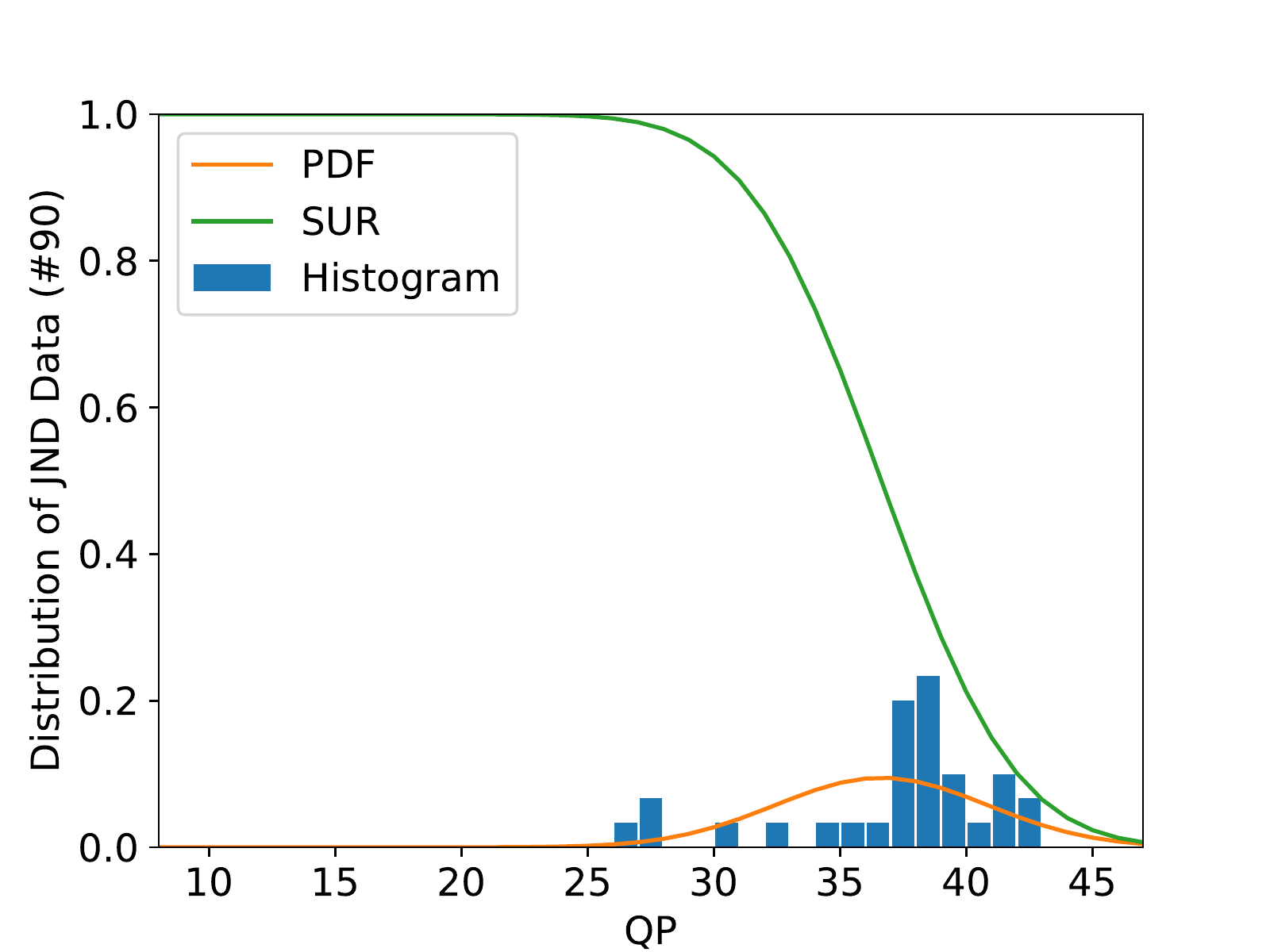}
\subcaption{}
\end{subfigure}
\begin{subfigure}[b]{0.3\linewidth}
\centering
\includegraphics[width=1.0\linewidth]{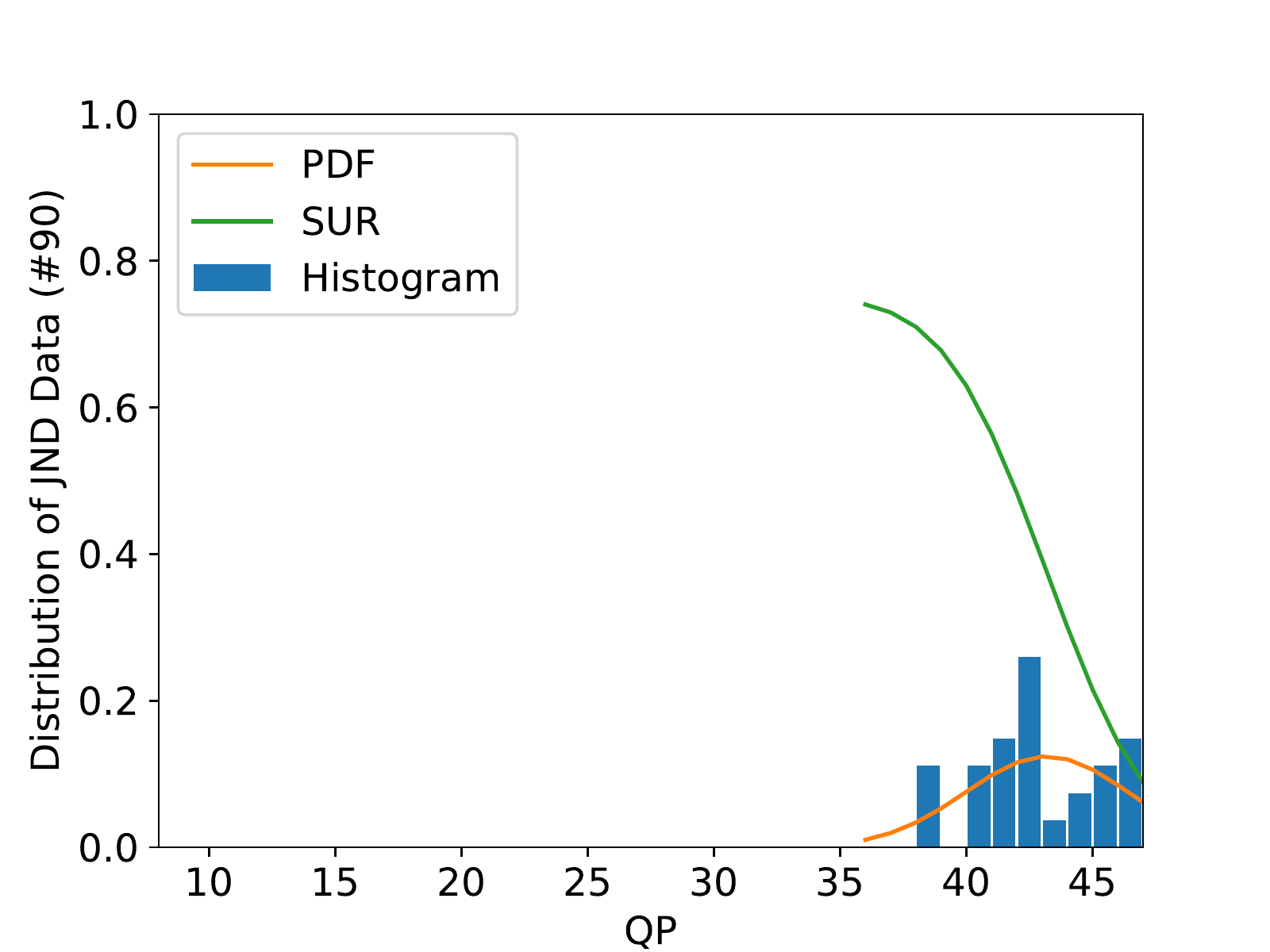}
\subcaption{}
\end{subfigure}
\begin{subfigure}[b]{0.3\linewidth}
\centering
\includegraphics[width=1.0\linewidth]{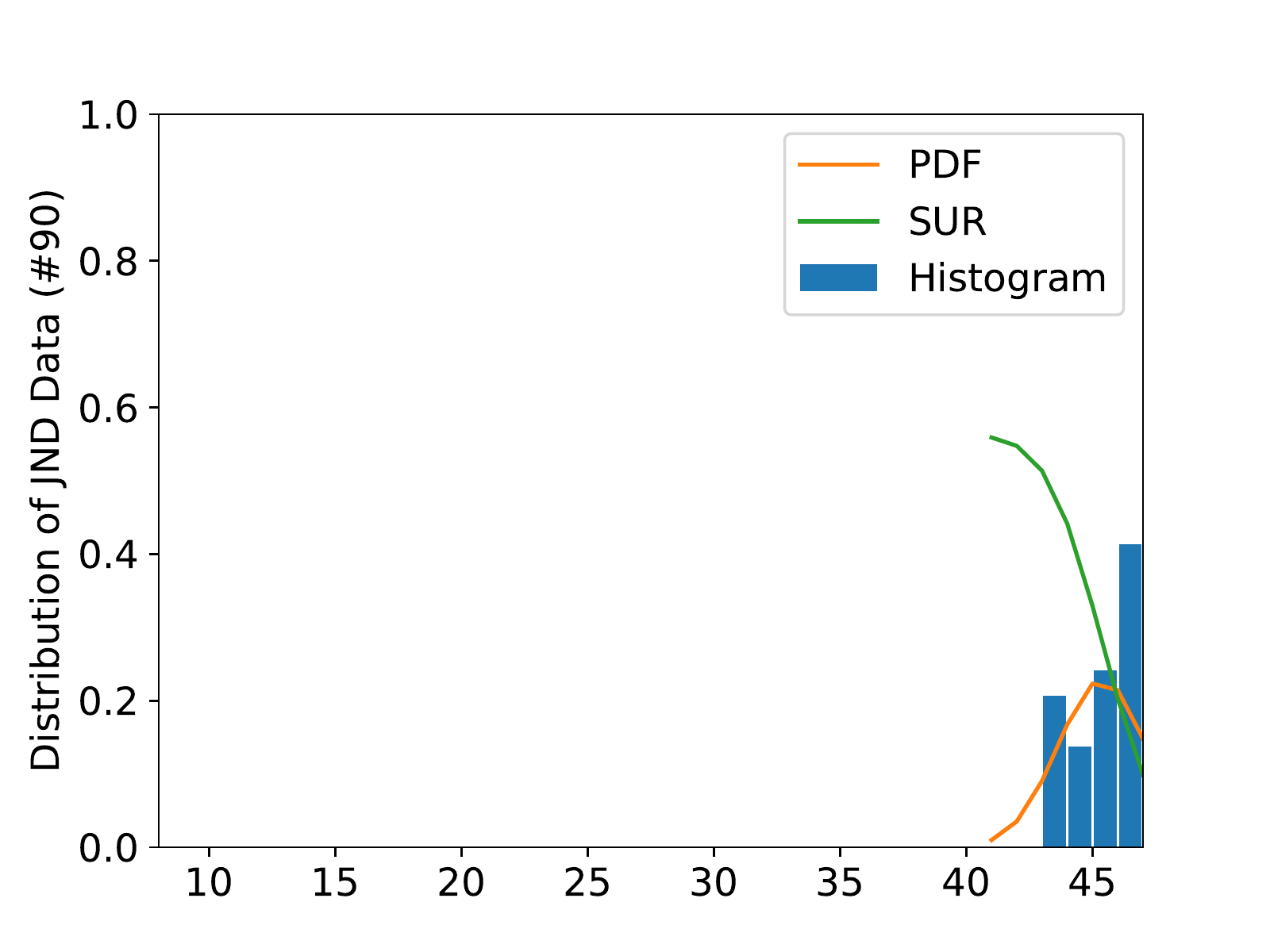}
\subcaption{}
\end{subfigure}
\caption{Histograms of collected 1st, 2nd and 3rd JND samples and their
PDF and SUR models are shown in the 1st, 2nd and 3rd columns,
respectively.  Figures in the 1st and 2nd rows are results from
sequences \#37 and \#90, respectively.}\label{fig:data_flow}
\end{figure*}

Fig. \ref{fig:frames_of_two_sequences} gives representative thumbnails
of two sequences, and their corresponding SUR curves for three JND
points are illustrated in Fig. \ref{fig:data_flow}. Taking sequence (a)
as an example to illustrate the modeling process of multiple JND points.
The reference for the first JND is perceptually lossless coded clip
$d_{0}$ and the location of the $75\%$ SUR point is $d_{18}$. Then, we
use $d_{18}$ as the reference for the second JND and get a bunch of JND
samples as the histogram of the second JND point, each of which is the
result of one test subject. The location of the $75\%$ SUR point for the
2nd JND is $d_{23}$. Finally, the third JND location can be derived
using $d_{23}$ as its reference for quality comparison.

\section{SUR Prediction for Multiple JND Points} \label{sec:sur_pred}

\begin{table}[!tb]
	\scriptsize
\centering
\caption{SUR and JND prediction settings. The main difference lies in
the reference used to predict the second and the third JND points.
\label{tab:settings}}
\begin{tabular}{{c}|{c}|{c}{c}{c}} \toprule
Models & Order & Reference & Samples  & JND \\ 
\cline{1-5}
                   & 1st         & $\mathbf{R}_{0}$    & $[1, 51]$    & $\mathbf{Y}_{1}$ \\ 
Subjective Test      & 2nd         & $\mathbf{Y}_{1}$    & [$\mathbf{Y}_{1}$+1, 51]    & $\mathbf{Y}_{2}$ \\ 
                   & 3rd         & $\mathbf{Y}_{2}$    & [$\mathbf{Y}_{2}$+1, 51]    & $\mathbf{Y}_{3}$ \\ 
\cline{1-5}
                 & 1st         & $\mathbf{R}_{0}$    & $[1, 51]$    & $\mathbf{\hat{Y}}_{1}$    \\
\cline{1-5}
Setting 1            & 2nd         & $\mathbf{Y}_{1}$    & [$\mathbf{Y}_{1}$+1, 51]    & $\mathbf{\hat{Y}}_{2}$   \\
Ground truth reference  & 3rd         & $\mathbf{Y}_{2}$    & [$\mathbf{Y}_{2}$+1, 51]    & $\mathbf{\hat{Y}}_{3}$   \\
\cline{1-5}
Setting 2         & 2nd         &$\mathbf{\hat{Y}}_{1}$    & [$\mathbf{\hat{Y}}_{1}$+1, 51]    & $\mathbf{\hat{Y}}_{2}$   \\
Predicted reference          & 3rd         & $\mathbf{\hat{Y}}_{2}$    & [$\mathbf{\hat{Y}}_{2}$+1, 51]    & $\mathbf{\hat{Y}}_{3}$  \\
\cline{1-5}
Setting 3              & 2nd         & $\mathbf{R}_{0}$    & $[1, 51]$    & $\mathbf{\hat{Y}}_{2}$   \\
Same reference         & 3rd         & $\mathbf{R}_{0}$    & $[1, 51]$    & $\mathbf{\hat{Y}}_{3}$    \\
\bottomrule
\end{tabular}
\end{table}

Apparently, the modeling JND/SUR curves exhibit differences in different
video sequences due to unique characteristics of each source content. To
predict the SUR curve, we mainly focus on the two factors in this work:
1) quality degradation due to compression, and 2) the masking effect due
to source content characteristics. To shed light on the impact of the
masking effect, we use sequences \#37 (DinnerTable) and \#90
(TodderFountain) as examples.  Their representative frames are shown in
Fig.  \ref{fig:frames_of_two_sequences} (a) and (b), and the histogram
of their first JND locations are given in Fig.  \ref{fig:data_flow} (a)
and (d), respectively. Sequence \#37 is a scene captured around a dining
table. It focuses on a male speaker with still dark background. The
speaker's face is the visual salient region that attracts people's
attention. The masking effect is weak and, as a result, the JND point
arrives earlier (i.e.  a smaller $j$ value in $d_j$). On the other hand,
sequence \#90 is a scene about a toddler playing in a fountain.  The
masking effect is strong due to water drops in background and fast
object movement.  As a result, compression artifacts are difficult to
perceive and the JND point arrives later.

The masking effect has significant impacts on the second and the third
JND. This phenomenon can be observed by comparing the second SUR curves
between sequence \#37 in Fig. \ref{fig:data_flow} (b) and sequence \#90
in Fig. \ref{fig:data_flow} (e), where \textit{DinnerTable} has much
smaller JND points than that of \textit{TodderFountain}. Indeed, even
the collected third JND points of \textit{DinnerTable} as shown in Fig.
\ref{fig:data_flow} (c) are smaller than that of the collected first JND
points of \textit{ToddlerFountain} as shown in Fig. \ref{fig:data_flow}
(d).  Thus, we need to pay special attention to the masking effect when
predicting the SUR curves.

A SUR prediction system was proposed in \cite{wang2017prediction}, which
focuses on the first JND. Here, we extend the framework further to the
second and third JND point prediction. We briefly review the prediction
method in \cite{wang2017prediction} for the purpose of completeness.
The quality of Spatial-Temporal Segments is
first evaluated by aggregating the similarity indices of local segments
to form a compact global index for each segment by using the VMAF metric \cite{vmaf}. Then, significant
segments are selected based on the slope of quality scores between
neighboring coded clips.  After that, we extract the masking effect that
reflects the unique characteristics of each video clip, and use the
support vector regression (SVR) to minimize the $L_{2}$ distance from
the SUR curves, and derive the JND point accordingly.

Since the JND point is reference dependent, we need to pin down the
reference clips in order to predict the second and the third JND points.
We evaluate the proposed system on different settings as shown in Table
\ref{tab:settings}. There are three different settings based on the
reference clips used to predict the second and the third JND.
\begin{itemize}
\item Setting 1: ground truth references without calibration errors.
\item Setting 2: predicted references from the previous SUR curve. It is
the practical scenario when no subjective data is available.
\item Setting 3: the same reference as the first JND point to verify the
robustness of the proposed method to the reference in a practical
prediction system.
\end{itemize}


\section{Experimental Results} \label{experiment_results}

\begin{table}[!tb]
	\scriptsize
\centering
\caption{Summary of averaged prediction errors of the first JND of video clips
in four resolutions.}
\label{tab:perf_comp_sur_1st}
\begin{tabular}{{c}|*{4}{c}} \toprule
              & 1080p & 720p & 540p  & 360p  \\
\cline{1-5}
$\Delta$SUR   & 0.039  & 0.038 & 0.037 & 0.042 \\
$\Delta$QP    & 1.218  & 1.273 & 1.345 & 1.605 \\
\bottomrule
\end{tabular}
\end{table}
In this section, we present the prediction results of the proposed SUR
prediction method on the VideoSet. It consists of 220 videos in 4
resolutions and three JND points per resolution per video clip. Here, we
focus on the SUR prediction of all the three JND points and conduct this
task for each video resolution independently. For each prediction task,
we train and test 220 video clips using the 5-fold validation. That is,
we choose 80\% (i.e. 176 video clips) as the training set and the
remaining 20\% (i.e., 44 video clips) as the testing set.
Since the JND location is chosen to be the QP value when the SUR value
is equal to 75\% in the VideoSet, we adopt the same rule here so that
the JND point can be easily computed from the predicted SUR curve.

The average prediction errors of the first SUR curve and the first JND
position for video clips in four resolutions are summarized in Table
\ref{tab:perf_comp_sur_1st}. Furthermore, Table
\ref{tab:perf_comp_sur_2nd} and Table \ref{tab:perf_comp_qp_2nd} show
$\Delta$SUR and $\Delta$QP for the second and the third JND under
different settings, respectively, where the $\Delta$SUR value is the sum
of absolute area differences between the predicted SUR curve and the
ground-truth SUR curve. We see that prediction errors increase as the resolution decreases for
the first JND point.  This is probably due to the use of fixed $W=320$
and $H=180$ values in generating spatial-temporal segments. The
parameter influence on prediction accuracy will be further investigated
in the future.  Moreover, the results in Table
\ref{tab:perf_comp_qp_2nd} show that Setting 1 always achieves the best
performance while Setting 3 gets the largest $\Delta$QP. However, we did
not observe the same phenomena in $\Delta$SUR in Table
\ref{tab:perf_comp_sur_2nd}. The reason is that the third JND point
(predicted) is the $75\%$ point on the curve which makes the remaining
QP range limited and SUR tends to be close to zero for large QP, e.g.,
$QP>40$.

\begin{figure*}[!th]
\centering
\begin{subfigure}[b]{0.3\linewidth}
\centering
\includegraphics[width=1.0\linewidth]{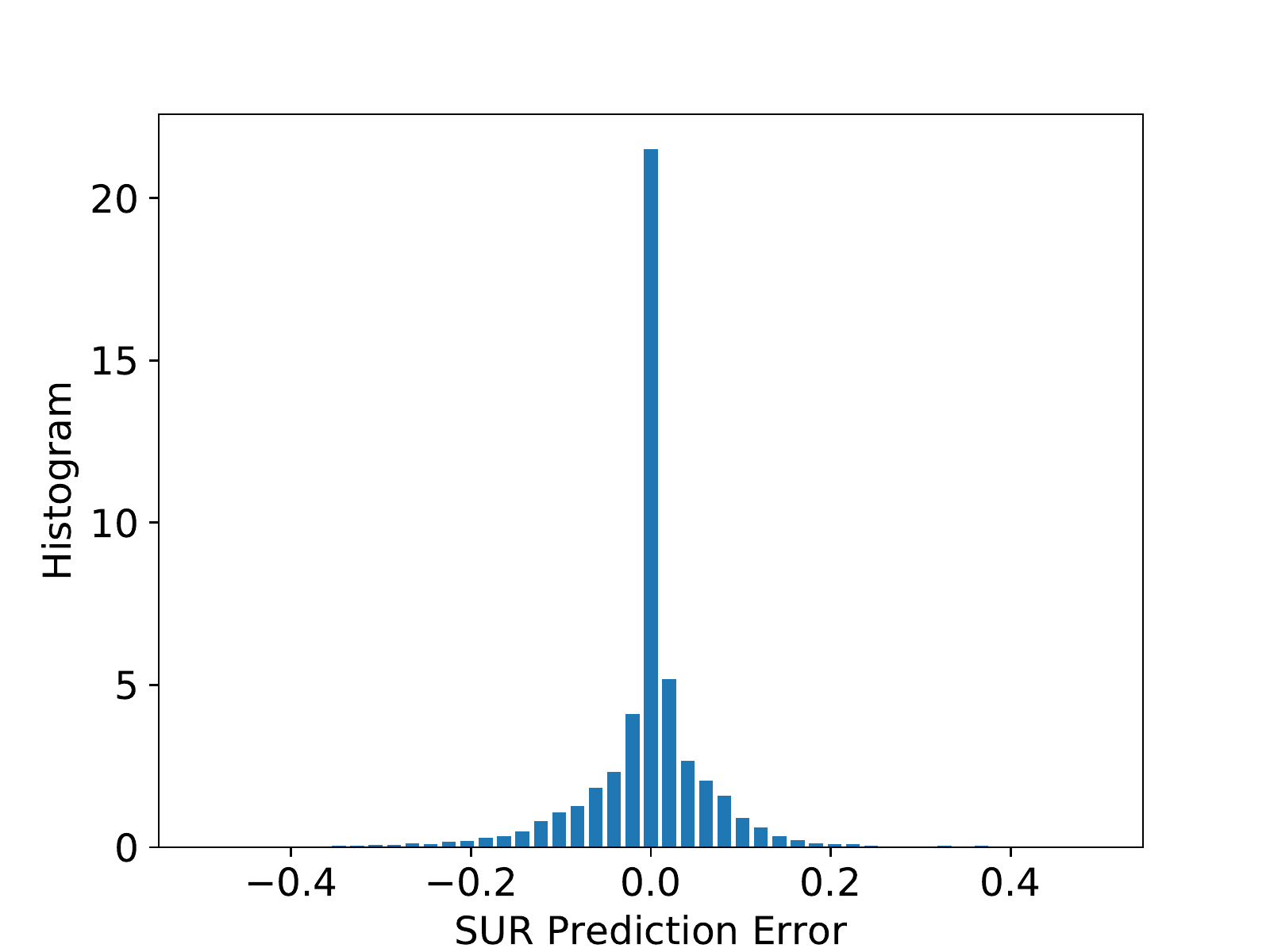}
\subcaption{}
\end{subfigure}
\begin{subfigure}[b]{0.3\linewidth}
\centering
\includegraphics[width=1.0\linewidth]{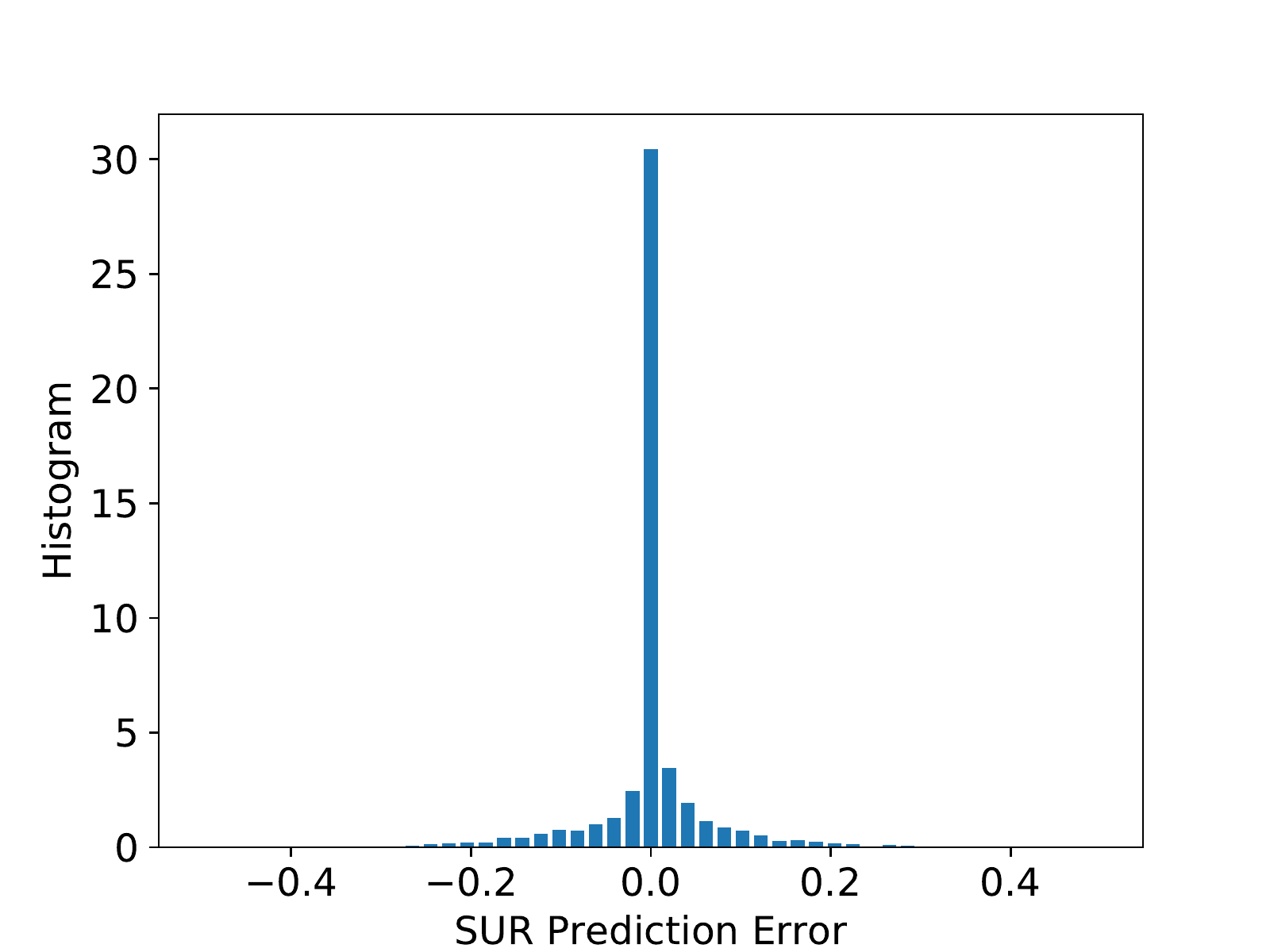}
\subcaption{}
\end{subfigure}
\begin{subfigure}[b]{0.3\linewidth}
\centering
\includegraphics[width=1.0\linewidth]{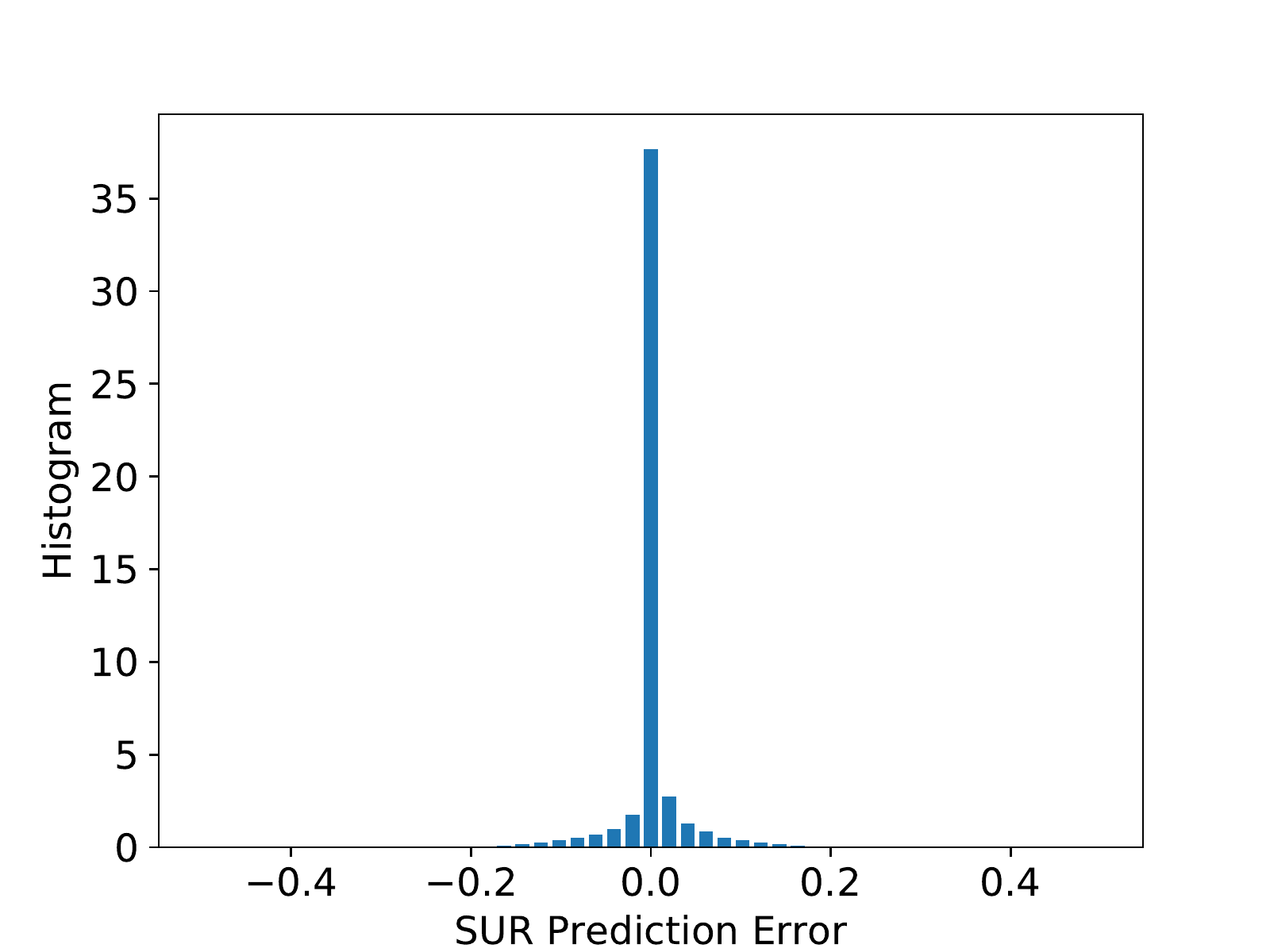}
\subcaption{}
\end{subfigure}
\begin{subfigure}[b]{0.3\linewidth}
\centering
\includegraphics[width=1.0\linewidth]{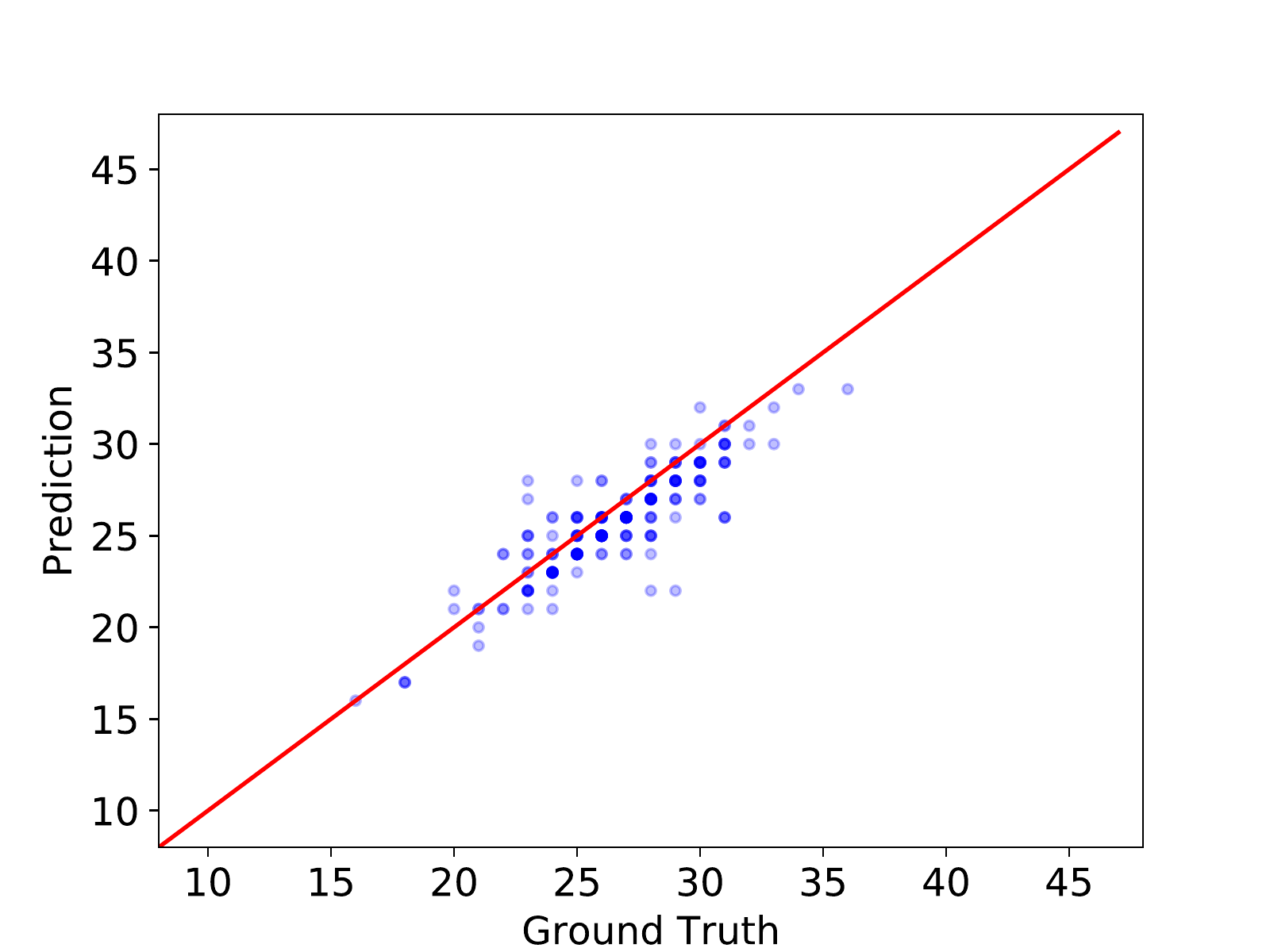}
\subcaption{}
\end{subfigure}
\begin{subfigure}[b]{0.3\linewidth}
\centering
\includegraphics[width=1.0\linewidth]{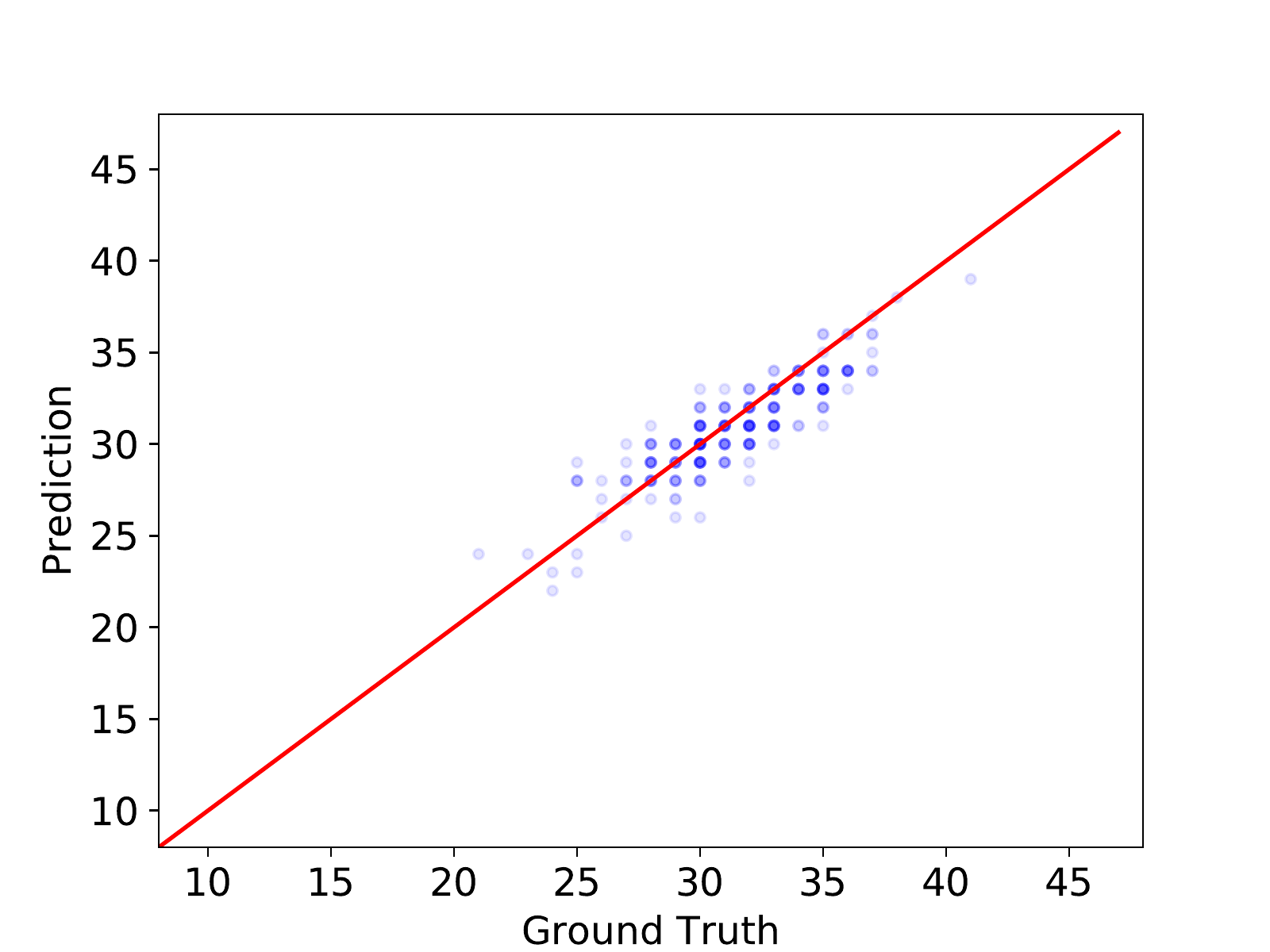}
\subcaption{}
\end{subfigure}
\begin{subfigure}[b]{0.3\linewidth}
\centering
\includegraphics[width=1.0\linewidth]{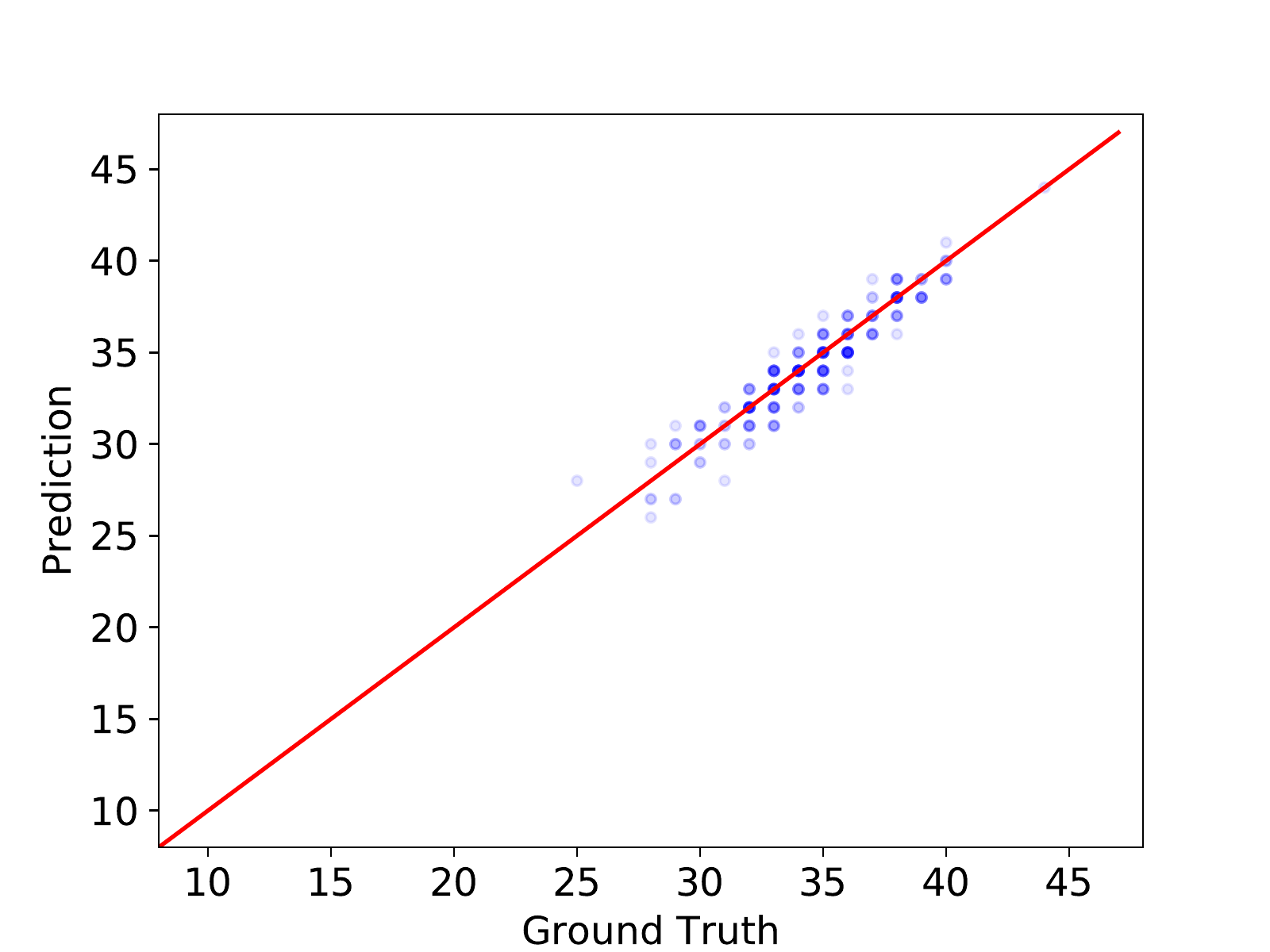}
\subcaption{}
\end{subfigure}
\caption{SUR/JND prediction results: histograms of $\Delta$SUR are shown
in (a)-(c) and the relationship between the predicted JND location and
the ground truth JND location are shown in (d)-(f). The left, middle and
right columns show results of the first, second and third JNDs,
respectively.}\label{fig:predicted_error}
\end{figure*}
We use 720p video as an example to demonstrate the prediction
performance of individual clips. The histograms of the SUR prediction
errors are shown in Fig. \ref{fig:predicted_error} (a)-(c) for the first
three SUR curves, where the mean absolute error (MAE) are 0.038, 0.054,
and 0.032, respectively. The predicted JND locations versus the
ground-truth JND locations are plotted in Fig.  \ref{fig:predicted_error}
(d)-(f), where each dot denotes one video clip.  As shown in the figure,
most dots are distributed along the 45-degree line, which indicates that
the predicted JND is very close to the ground truth JND for most
sequences.

\begin{table}[!tb]
	\scriptsize
\centering
\caption{Mean Absolute Error (MAE) of predicted SUR, i.e., $\Delta$SUR
for the second and the third JND. \label{tab:perf_comp_sur_2nd}}
\begin{tabular}{{c}|*{2}{c}|*{2}{c}|*{2}{c}} \toprule
 & \multicolumn{2}{c}{Setting 1} & \multicolumn{2}{c}{Setting 2} & \multicolumn{2}{c}{Setting 3} \\
Resolution  & 2nd & 3rd  & 2nd & 3rd & 2nd & 3rd \\
\cline{1-7}
1080p                & 0.062 & 0.029 & 0.063 & 0.065 & 0.057 & 0.056 \\
720p                 & 0.054 & 0.032 & 0.057 & 0.060 & 0.055 & 0.056 \\
540p                 & 0.050 & 0.030 & 0.054 & 0.052 & 0.046 & 0.049 \\
360p                 & 0.052 & 0.030 & 0.058 & 0.056 & 0.048 & 0.053 \\
\bottomrule
\end{tabular}
\end{table}
\begin{table}[!tb]
	\scriptsize
\centering
\caption{Mean Absolute Error (MAE) of predicted JND location, i.e.,
$\Delta$QP for the second and the third JND
\label{tab:perf_comp_qp_2nd}}
\begin{tabular}{{c}|*{2}{c}|*{2}{c}|*{2}{c}} \toprule
 & \multicolumn{2}{c}{Setting 1} & \multicolumn{2}{c}{Setting 2} & \multicolumn{2}{c}{Setting 3} \\
Resolution  & 2nd & 3rd  & 2nd & 3rd & 2nd & 3rd \\
\cline{1-7}
1080p               & 1.618 & 0.709 & 2.009 & 2.245 & 2.364 & 2.445 \\
720p                & 1.227 & 0.750 & 1.709 & 1.927 & 2.209 & 2.227 \\
540p                & 1.223 & 0.773 & 1.523 & 1.700 & 1.873 & 2.009 \\
360p                & 1.341 & 0.745 & 1.750 & 1.836 & 1.923 & 2.105 \\
\bottomrule
\end{tabular}
\end{table}

\section{Conclusion and Future Work}\label{sec:conclusion}

A JND-based video quality model was proposed in this paper. The proposed
video quality index, \textit{i.e.} the SUR, seamlessly
reflects the perceived quality of compressed video clips. We presented a
SUR prediction framework for the first three JND
points and the efficiency of the proposed SUR prediction method is
verified on the VideoSet in different settings, especially for the
second and the third JND points. The proposed method achieves good
performance in all video sequences with different resolutions. In the
future, we would like to investigate the influence of different
parameters in the proposed method, \textit{e.g.,} the dimension of
segment, percentage of key segments and other sophisticated
spatial-temporal pooling method, to improve the performance of the JND
point prediction.

\bibliographystyle{IEEEtran}
\bibliography{references,refs}

\begin{thebibliography}{10}
\providecommand{\url}[1]{#1}
\csname url@samestyle\endcsname
\providecommand{\newblock}{\relax}
\providecommand{\bibinfo}[2]{#2}
\providecommand{\BIBentrySTDinterwordspacing}{\spaceskip=0pt\relax}
\providecommand{\BIBentryALTinterwordstretchfactor}{4}
\providecommand{\BIBentryALTinterwordspacing}{\spaceskip=\fontdimen2\font plus
\BIBentryALTinterwordstretchfactor\fontdimen3\font minus
  \fontdimen4\font\relax}
\providecommand{\BIBforeignlanguage}[2]{{%
\expandafter\ifx\csname l@#1\endcsname\relax
\typeout{** WARNING: IEEEtran.bst: No hyphenation pattern has been}%
\typeout{** loaded for the language `#1'. Using the pattern for}%
\typeout{** the default language instead.}%
\else
\language=\csname l@#1\endcsname
\fi
#2}}
\providecommand{\BIBdecl}{\relax}
\BIBdecl

\bibitem{sandvine2016global}
I.~SANDVINE, ``Global internet phenomena report,'' 2016.

\bibitem{per_title_encode}
Netflix, ``Per-title encode optimization,'' \emph{The Netflix Tech Blog}, 2015.

\bibitem{ssim}
Z.~Wang, A.~C. Bovik, H.~R. Sheikh, and E.~P. Simoncelli, ``Image quality
  assessment: from error visibility to structural similarity,'' \emph{IEEE
  Transactions on Image Processing}, vol.~13, no.~4, pp. 600--612, 2004.

\bibitem{fsim}
L.~Zhang, L.~Zhang, X.~Mou, and D.~Zhang, ``Fsim: A feature similarity index
  for image quality assessment,'' \emph{IEEE Transactions on Image Processing},
  vol.~20, no.~8, pp. 2378--2386, 2011.

\bibitem{dlm}
S.~Li, F.~Zhang, L.~Ma, and K.~N. Ngan, ``Image quality assessment by
  separately evaluating detail losses and additive impairments,'' \emph{IEEE
  Transactions on Multimedia}, vol.~13, no.~5, pp. 935--949, 2011.

\bibitem{lin2015experimental}
J.~Y. Lin, L.~Jin, S.~Hu, I.~Katsavounidis, Z.~Li, A.~Aaron, and C.-C.~J. Kuo,
  ``Experimental design and analysis of {JND} test on coded image/video,'' in
  \emph{SPIE Optical Engineering+ Applications}.\hskip 1em plus 0.5em minus
  0.4em\relax International Society for Optics and Photonics, 2015, pp.
  95\,990Z--95\,990Z.

\bibitem{jin2016statistical}
L.~Jin, J.~Y. Lin, S.~Hu, H.~Wang, P.~Wang, I.~Katsavounidis, A.~Aaron, and
  C.-C.~J. Kuo, ``Statistical study on perceived {JPEG} image quality via
  {MCL-JCI} dataset construction and analysis,'' \emph{Electronic Imaging},
  vol. 2016, no.~13, pp. 1--9, 2016.

\bibitem{mcl_jcv}
H.~Wang, W.~Gan, S.~Hu, J.~Y. Lin, L.~Jin, L.~Song, P.~Wang, I.~Katsavounidis,
  A.~Aaron, and C.-C.~J. Kuo, ``{MCL-JCV}: a {JND}-based {H.264/AVC} video
  quality assessment dataset,'' in \emph{Image Processing (ICIP), 2016 IEEE
  International Conference on}.\hskip 1em plus 0.5em minus 0.4em\relax IEEE,
  2016, pp. 1509--1513.

\bibitem{Wang2017292}
H.~Wang, I.~Katsavounidis, J.~Zhou, J.~Park, S.~Lei, X.~Zhou, M.-O. Pun,
  X.~Jin, R.~Wang, X.~Wang \emph{et~al.}, ``{VideoSet}: A large-scale
  compressed video quality dataset based on {JND} measurement,'' \emph{Journal
  of Visual Communication and Image Representation}, no.~46, pp. 292--302,
  2017.

\bibitem{aimar2005x264}
L.~Aimar, L.~Merritt, E.~Petit, M.~Chen, J.~Clay, M.~Rullgrd, C.~Heine, and
  A.~Izvorski, ``X264-a free {H.264/AVC} encoder,''
  \url{http://www.videolan.org/developers/x264.html}, 2005, accessed: 04/01/07.

\bibitem{huang2017measure}
Q.~Huang, H.~Wang, S.~C. Lim, H.~Y. Kim, S.~Y. Jeong, and C.-C.~J. Kuo,
  ``Measure and prediction of {HEVC} perceptually lossy/lossless boundary {QP}
  values,'' in \emph{Data Compression Conference (DCC), 2017}.\hskip 1em plus
  0.5em minus 0.4em\relax IEEE, 2017, pp. 42--51.

\bibitem{wang2017prediction}
H.~Wang, I.~Katsavounidis, Q.~Huang, X.~Zhou, and C.-C.~J. Kuo, ``Prediction of
  satisfied user ratio for compressed video,'' \emph{arXiv preprint
  arXiv:1710.11090}, 2017.

\bibitem{jarque1987test}
C.~M. Jarque and A.~K. Bera, ``A test for normality of observations and
  regression residuals,'' \emph{International Statistical Review/Revue
  Internationale de Statistique}, pp. 163--172, 1987.

\bibitem{vmaf}
Z.~Li, A.~Aaron, I.~Katsavounidis, A.~Moorthy, and M.~Manohara, ``Toward a
  practical perceptual video quality metric,'' \emph{The Netflix Tech Blog},
  vol.~6, 2016.

\end{thebibliography}

\end{document}